\documentclass[10pt,preprint]{aastex}
\title {Disentangling Morphology, Star Formation, Stellar Mass, and Environment in Galaxy Evolution}
\author {Daniel Christlein\altaffilmark{1,2} and Ann I. Zabludoff\altaffilmark{1}}
\altaffiltext{1}{Steward Observatory, The University of Arizona}
\altaffiltext{2}{present affiliation: Universidad de Chile \& Yale University}
\email {christlein@astro.yale.edu \\ azabludoff@as.arizona.edu}
\date {\today}
\begin {document}
\begin {titlepage}
\shortauthors{Christlein, \& Zabludoff}
\shorttitle{Disentangling Galaxy Evolution} 
\begin {abstract}

We present a study of the spectroscopic and photometric properties of galaxies in six nearby clusters. We perform a partial correlation analysis on our dataset to investigate whether the correlation between star formation rates in galaxies and their environment is merely another aspect of correlations of morphology, stellar mass, or mean stellar age with environment, or whether star formation rates vary independently of these other correlations. We find a residual correlation of ongoing star formation with environment, indicating that even galaxies with similar morphologies, stellar masses, and mean stellar ages have lower star formation rates in denser environments. Thus, the current star formation gradient in clusters is not just another aspect of the morphology-density, stellar mass-density, or mean stellar age-density relations. Furthermore, the star formation gradient cannot be solely the result of initial conditions, but must partly be due to subsequent evolution through a mechanism (or mechanisms) sensitive to environment. Our results constitute a true ``smoking gun'' pointing to the effect of environment on the later evolution of galaxies.
\end{abstract}
\end{titlepage}
\section{Introduction}

Galaxies exhibit a great diversity in mass, luminosity, morphology, and star formation activity. It is now generally recognized that these properties are correlated with the environment wherein a given galaxy resides. Such correlations are expected in hierarchical models of galaxy evolution, which predict the oldest, most massive, least star-forming galaxies to reside in the densest regions of the universe. Subsequent evolution brings less massive, still star-forming galaxies into these denser regions from the field. This leads us to expect gradients in galaxy properties such as morphology and star formation with environment, even from initial conditions alone. If galaxies are affected by environmentally-dependent mechanisms later in their lives, these gradients may be altered.

Observationally, one of the best-established among these correlations is the morphology-environment relation \citep{dressler80}, in which the morphological composition of the galaxy population is shifted towards earlier types --- S0s and ellipticals --- in denser environments such as groups or clusters of galaxies, near the center of a cluster, or in other regions of locally enhanced galaxy number density. 

Then there is the correlation between star formation and environment: galaxies have lower star formation rates in dense environments than elsewhere. This correlation has also been known for a long time \citep{gisler78} and recently confirmed by large modern surveys \citep{lewis,gomez}, which demonstrate that the star formation rates in galaxies near clusters vary as a function of the distance of the galaxy from the cluster center. Such results have been interpreted as evidence for an effect of the cluster environment on star formation. However, these studies do not provide unambiguous proof for such an effect, because three important questions remain:

1) Is the star formation gradient in clusters simply a consequence of the morphology-environment relation? Generally, early-type galaxies have lower star formation rates than late-types (as the result of initial conditions and/or subsequent evolution).  As a consequence, we would expect a star formation gradient in clusters.  Is this gradient as strong as that observed? Or is the evolution of star formation rates in different local environments, at least to some extent, independent of the morphological evolution, either because different mechanisms act in different environments, or because the same mechanism affects star formation and morphology differently in different environments?

2) Is the star formation gradient simply a consequence of variations in the galaxy luminosity or mass function with environment? Some evidence \citep{depropris} exists that the faint end of the luminosity function is shallower in the cores of clusters than on the outskirts. Since low-mass, faint galaxies are typically more strongly star-forming, the lower rates of star formation in clusters could be due to the fact that samples of dense environments contain a larger fraction of giant galaxies. As long as this question is unresolved, it is not clear whether environment has a direct effect on the star formation rates of individual galaxies, or whether the star formation gradient is simply a result of variations of the luminosity or mass function with environment (which could be due to initial conditions or subsequent evolution).

3) Does the star formation gradient in clusters simply arise because clusters contain a large fraction of long-quiescent galaxies? Although individual objects are known that could be in the process of being transformed by mergers (e.g., Yang et al. 2004) or ram pressure stripping (e.g. Vollmer et al. 2004a, 2004b), this does not constitute evidence that such ongoing transformations are primarily responsible for the gradients in morphology and star formation with environment. Many galaxies in clusters contain very old stellar populations and may have been quiescent for many Gyr, possibly since their formation. The star formation gradient could simply reflect initial conditions: these old galaxies constitute an increasing fraction of all galaxies towards the cluster center, as star-forming galaxies from the field dominate the cluster outskirts. Claiming an ongoing effect of environment on star formation would require us to to detect a star formation gradient even for galaxies with comparable star formation histories (i.e., mean stellar ages).

Testing whether dense environments have a direct and ongoing effect on star formation rates in clusters therefore requires us to examine whether an independent correlation of star formation with environment exists that cannot be accounted for by the correlations of morphology, stellar age, and stellar mass with environment. Such an analysis must be based on measurements of galaxy spectra and structural parameters in a large sample. The observational capabilities for such a program, particularly for multi-fiber spectroscopy of a large number of objects, as well as the computational capabilities to quantitatively characterize the morphologies of a large sample of galaxies, have not existed until recently. A few studies \citep{gomez,balogh98,dressler99,poggianti99} have attempted to address this question by splitting galaxies into broad morphological classes and examining the dependence of star formation within each class on environment. This procedure suffers from the problem that a coarse binning in morphology may leave residual correlations between star formation and morphology within each bin, and it is therefore not clear whether the residual correlations reported in these studies are really independent of the morphology-environment relation. \citet{hashimoto} and \citet{koopmann98} use the concentration index as a quantitative measure of morphology, which allows for a finer and more reproducible binning in morphology. 

Generally, these past studies have favored the hypothesis that the correlation of star formation with environment cannot be explained by the morphology-density relation alone, i.e., that morphology and star formation are at least partially decoupled. This conclusion is supported by the observation in \citet{kauffmann2004} that the star formation-environment relation is the strongest correlation of galaxy properties with environment. Kauffmann et al. also demonstrate that the correlation between stellar mass and environment alone cannot account for the star formation gradient. However, none of these studies provide a constraint on whether the lower star formation rate measured in denser environments is due to initial conditions or more recent environmental effects, nor do they address the question of whether the {\it combined} variations of morphology, stellar mass, and mean stellar age with environment may be strong enough to explain the star formation gradient.

In this paper, we present the results of a multi-variate analysis of a spectroscopic sample of galaxies in six nearby clusters. We use a partial correlation analysis to examine whether there is a residual correlation between star formation and environment even when holding morphology, stellar mass, and mean stellar age fixed. Using partial correlation coefficients allows us to account for a large number of variables simultaneously and thus to disentangle the effect of environment from all other correlations, in order to isolate any ongoing, direct impact of environment on star formation. With this approach, we also avoid the problems associated with the usual method of representing morphology by coarse binning. 

This paper is organized as follows: In \S \ref{sec_data} we briefly review the sample upon which our analysis is based. In \S \ref{sec_control}, we discuss the variables --- star formation, morphology, stellar mass, mean stellar age --- that we use in our search for residual correlations between star formation and environment. We represent ongoing star formation with the equivalent width of the [OII] $\lambda 3727$ doublet, and recent star formation with the equivalent width of H$\delta$ in absorption in the absence of [OII] emission. We characterize morphology with the bulge fraction, stellar mass with a combination of near-IR photometry and the {4000\AA} break strength, and mean stellar age with the {4000\AA} break strength. In \S \ref{sec_analysis}, we review the mathematical tools used for the analysis. We discuss how we calculate partial correlation coefficients to address the question of whether star formation varies differently with environment than morphology, stellar mass, or mean stellar age. We also present our procedure for calculating completeness corrections using the Discrete Maximum Likelihood method \citep{cmz2003,cz2004} and discuss the limits of our survey and its completeness within these limits. In \S \ref{sec_results}, we present the results of our analysis. In \S \ref{sec_morden}, we explore different measures of environment and choose the projected radial distance of a galaxy from the cluster center as our environmental variable. We then verify that we can reproduce the morphology-environment and star formation-environment relations. In \S \ref{sec_residual}, we examine whether residual correlations of current and recent star formation with environment remain when accounting for the effects of morphology, stellar mass, and mean stellar age, and quantify the magnitude of the residual star formation gradient relative to other effects. In \S \ref{sec_errors}, we estimate the effect that observational uncertainties in our determinations of morphology, stellar mass, and mean stellar age have on the residual star formation gradient. In \S \ref{sec_conclusions}, we present our conclusions.

\section{THE CLUSTER SURVEY}
\label{sec_data}
Our sample consists of cluster galaxies from a spectroscopic and $R$-band imaging survey of six nearby clusters. The spectroscopic survey ensures that contamination of the sample by background field galaxies is minimized. Table \ref{tabclusters} lists some parameters of these six clusters for $H_{0}=71$ km s$^{-1}$ Mpc$^{-1}$, $\Omega_{m}=0.3$ and $\Omega_{\Lambda}=0.7$, as applied throughout this paper. For details regarding the survey and data reduction, we refer readers to \citet{cz2003}. 

\begin{deluxetable}{lrrrccrc}
\tabletypesize{\scriptsize}
\tablecaption{The Cluster Sample}
\tablewidth{0pt}
\tablehead{
\colhead{Cluster} & \colhead{centroid} &\colhead{N}&\colhead{$\overline{cz}$}&\colhead{$\Delta m$}&\colhead{$\Delta cz$}&\colhead{$\sigma$}&\colhead{$r_{sampling}$}   \\
                  & \colhead{(RA, Dec; J2000)}  &           & \colhead{[km/s]}        &\colhead{[mag]}                  &\colhead{[km/s]}    &\colhead{[km/s]}  &\colhead{[Mpc]}  
}
\startdata
A1060&10 36 51.29 -27 31 35.3&252&$3683\pm46$&33.59&2292  - 5723&$724\pm31$&0.67   \\
A496 &04 33 37.09 -13 14 46.3&241&$9910\pm48$&35.78&7731  - 11728&$728\pm36$&1.76  \\
A1631&12 52 49.84 -15 26 17.1&340&$13844\pm39$&36.53&12179 - 15909&$708\pm28$&2.42 \\
A754 &09 08 50.08 -09 38 11.8&415&$16369\pm47$&36.90&13362 - 18942&$953\pm40$&2.83 \\
A85  &00 41 37.81 -09 20 33.2&280&$16607\pm60$&36.94&13423 - 19737&$993\pm53$&2.87 \\
A3266&04 31 11.92 -61 24 22.7&331&$17857\pm69$&37.10&14129 - 21460&$1255\pm58$&3.07\\
\enddata
\tablecomments{$N$ is the number of sampled galaxies per cluster. $\overline{cz}$ is the mean velocity, $\Delta m$ the distance modulus (for $H_{0}=71$ km s$^{-1}$ Mpc$^{-1}$). $\Delta cz$ is the velocity range spanned by cluster members, $\sigma$ is the line-of-sight velocity dispersion, and $r_{sampling}$ is the projected physical radius sampled (centroid-to-edge).} 
\label{tabclusters}
\end{deluxetable}

\section{GALAXY PROPERTIES}
\label{sec_control}

\subsection{Star Formation Indices}
\label{sec_sfindices}

Because the spectra for our sample are not flux calibrated, it is not possible to calculate absolute line fluxes and star formation rates. Instead, we use the equivalent width of the $[OII]\lambda 3727$ doublet as a measure of current star formation rate relative to the existing blue continuum luminosity. Although sensitive to dust and metallicity effects \citep{kewley}, $EW([OII])$ is a sufficiently accurate indicator of star formation rate \citep{kennicutt} for our purposes, which only require an approximate relative measure, not a precise calibration. Contamination of the sample with AGN is not a major concern: only one of the cluster members in our survey is currently known to be a Seyfert galaxy (NASA/IPAC Extragalactic Database; Helou et al. 1991). We discuss possible systematic errors introduced by using the OII EW in \S \ref{sec_residualcurrent}.

A suitable measure of recent, but not on-going, star formation is the H$\delta$ Balmer absorption line in the absence of significant [OII] emission. Balmer absorption lines are typically strongest in galaxies whose light is dominated by stars of spectral type A, and so their equivalent widths are particularly sensitive to a galaxy's star formation activity over the timescale of the past $\sim$ Gyr. A caveat is that, in strongly star-forming galaxies, Balmer emission from HII regions can fill the Balmer absorption lines and thus affect our measurement of the $H\delta$ equivalent width. To avoid this, we only consider galaxies that have no significant [OII] emission (defined as having [OII] line emission detectable at less than 2$\sigma$; see Zabludoff et al. 1996). 

To calculate an equivalent width, the local continua are fit over the 100 pixels ($\sim 250$\AA) on either side of the line that exclude the line itself  and the nearby sky lines. Beginning at the line center, the line is integrated outward until reaching the continuum level. That uncalibrated flux and the interpolated value of the continuum at line center are used to calculate the equivalent width. The equivalent width uncertainties, which are typically less than 1 \AA, are calculated using counting statistics (the detector is a photon counter with approximately zero read noise), the local noise in the continuum, and standard propagation of errors. Equivalent widths are cosmologically corrected.

Values of $EW([OII])$ and $EW(H\delta)$ for our sample are tabulated in Table \ref{tab_spec}.

\begin{deluxetable}{lrrrrrrrr}
\rotate
\tabletypesize{\scriptsize}
\tablecaption{Spectroscopic Catalog (Example)}
\tablewidth{0pt}
\tablehead{
\colhead{ID} & \colhead{RA (J2000)} & \colhead{Dec (J2000)} & \colhead{$D4000$} & \colhead{$\Delta D4000$} & \colhead{$EW([OII])$} & \colhead{$\Delta EW([OII])$} & \colhead{$EW(H\delta)$} & \colhead{$\Delta EW(H\delta)$} \\
\colhead{} & \colhead{} & \colhead{} & \colhead{} & \colhead{} & \colhead{[\AA]} & \colhead{[\AA]} & \colhead{[\AA]} & \colhead{[\AA]}
}
\startdata
   1060A\_295[11] & 10 36 13.46 & -27 31 54.90 &   1.828 &   0.073 &    1.80 &    2.10 &   -0.10 &    0.57 \\
   1060A\_295[26] & 10 35 37.40 & -27 10 49.60 &   1.893 &   0.103 &    3.13 &    2.87 &    1.53 &    1.34 \\
   1060A\_295[41] & 10 34 48.52 & -26 53 30.40 &   1.604 &   0.075 &    2.65 &    2.21 &    2.02 &    1.22 \\
    1060A\_393[1] & 10 36 49.04 & -27 23 18.60 &   2.257 &   0.057 &   -0.16 &    0.56 &    0.30 &    0.59 \\
    1060A\_393[4] & 10 36 35.60 & -27 08 56.20 &   1.852 &   0.050 &   -0.28 &    0.63 &    0.07 &    0.42 \\
    1060A\_393[6] & 10 36 26.76 & -27 23 24.80 &   1.741 &   0.056 &    0.48 &    1.24 &    0.08 &    0.97 \\
    1060A\_393[7] & 10 35 55.70 & -27 14 11.90 &   1.997 &   0.048 &    0.72 &    0.97 &    1.16 &    1.01 \\
   1060A\_393[10] & 10 36 19.30 & -27 28 45.90 &   1.786 &   0.055 &   -1.28 &    0.74 &    0.69 &    0.67 \\
   1060A\_393[12] & 10 35 46.79 & -27 38 48.70 &   1.515 &   0.037 &    8.43 &    1.22 &    2.76 &    1.14 \\
   1060A\_393[13] & 10 35 38.27 & -27 31 34.30 &   1.892 &   0.034 &    0.99 &    0.79 &    1.21 &    0.94 \\
   1060A\_393[14] & 10 35 21.68 & -27 23 26.20 &   1.365 &   0.026 &    4.44 &    0.77 &    7.42 &    1.32 \\
   1060A\_393[21] & 10 36 29.47 & -27 45 29.10 &   1.727 &   0.046 &    0.04 &    0.65 &    1.32 &    0.61 \\
   1060A\_393[22] & 10 36 21.38 & -27 46 30.60 &   1.721 &   0.052 &    0.92 &    1.60 &    1.96 &    0.96 \\
   1060A\_393[23] & 10 36 03.81 & -27 55 08.30 &   1.787 &   0.047 &   -0.26 &    0.65 &    0.79 &    0.61 \\
   1060A\_393[24] & 10 36 01.81 & -27 41 07.00 &   1.955 &   0.039 &    0.68 &    0.81 &    0.33 &    0.51 \\
   1060A\_393[25] & 10 35 55.21 & -27 45 16.60 &   2.110 &   0.053 &    3.91 &    1.88 &    1.50 &    1.04 \\
   1060A\_393[27] & 10 35 53.88 & -27 22 19.90 &   2.126 &   0.053 &   -0.70 &    0.60 &   -0.07 &    0.39 \\
   1060A\_393[35] & 10 34 56.05 & -27 38 24.60 &   1.396 &   0.037 &   14.06 &    1.48 &    2.38 &    1.01 \\
   1060A\_393[38] & 10 34 26.69 & -27 30 03.50 &   1.343 &   0.028 &   51.28 &    1.46 &    0.26 &    0.60 \\
   1060A\_393[41] & 10 34 58.90 & -26 52 28.70 &   1.787 &   0.057 &   -0.65 &    0.75 &    1.24 &    1.08 \\
   1060A\_393[42] & 10 34 45.84 & -26 57 17.90 &   1.709 &   0.055 &   -0.29 &    0.74 &   -0.08 &    0.74 \\
   1060A\_393[43] & 10 34 23.74 & -26 59 54.30 &   2.459 &   0.056 &    0.38 &    1.22 &    1.09 &    0.90 \\
\enddata
\tablecomments{The full table will be published in the electronic version of the Astrophysical Journal and can be provided upon request by the authors. The $R$-band magnitudes are published in \citet{cz2003}.}
\label{tab_spec}
\end{deluxetable}

\subsection{Morphologies}
\label{sec_bd}
We use the GIM2D software \citep{gim2d} to perform a two-dimensional decomposition of the galaxy images into a bulge component, described by a de Vaucouleurs surface brightness profile \citep{devauc}, and a disk component with an exponential surface brightness profile. We then calculate the fraction $B/T$ of the total luminosity associated with the bulge. Prior to the fit, we transform all galaxy images to a fiducial rest frame $cz=17858$ $km/s$ (corresponding to the mean velocity of the most distant cluster, A3266) by fading the surface brightnesses by $(z_{cosmological}+1)^{-2}(z_{total}+1)^{-2}$, smearing the images to achieve a consistent FWHM of 2 arcsec, and rebinning their pixels with the new angular diameter distance to achieve the same physical resolution per pixel. This approach ensures that determinations of $B/T$ are internally consistent among the clusters, which span a mean velocity range from 3682 to 17858 km/s. The final catalog contains bulge-disk decompositions of 1637 galaxies (1304 of them for galaxies with $M_{R}\leq-19.2$). This procedure is also discussed in \citet{cz2004}.

The limit of $m_{R}=18$ that we adopt for the spectroscopic catalog as well as for the bulge-disk decompositions corresponds to $M_{R}\approx-19.2$ in the fiducial field. For fainter galaxies, the signal-to-noise ratio is too low to permit reliable fits, and spectroscopic information is not available for all clusters. We therefore use $M_{R}=-19.2$ as the magnitude limit for all our analyses in \S \ref{sec_results}. 

Table \ref{tab_catalog} contains the coordinates and $B/T$ values used in this paper.

\begin{deluxetable}{lrrrrrrrrr}
\rotate
\tabletypesize{\scriptsize}
\tablecaption{Morphological Catalog (Example)}
\tablewidth{0pt}
\tablehead{
\colhead{ID} &\colhead{RA (J2000)} &\colhead{Dec (J2000)} & \colhead{$B/T$} & \colhead{$B/T_{min}$} & \colhead{$B/T_{max}$} & \colhead{$B/T_{f}$} & \colhead{$B/T_{f,min}$} & \colhead{$B/T_{f,max}$} & \colhead{$\omega$} 
}
\startdata
   1060A\_295[11] & 10 36 13.46 & -27 31 54.90 & 0.03 & 0.00 & 0.14 & 0.82 & 0.00 & 1.00 &   4.872 \\
   1060A\_295[26] & 10 35 37.40 & -27 10 49.60 & 0.13 & 0.00 & 0.25 & 0.40 & 0.00 & 1.00 &   5.690 \\
   1060A\_295[41] & 10 34 48.52 & -26 53 30.40 & 0.00 & 0.00 & 0.09 & 0.78 & 0.00 & 1.00 &  10.319 \\
    1060A\_393[1] & 10 36 49.04 & -27 23 18.60 & 0.68 & 0.64 & 0.73 & 0.60 & 0.12 & 1.00 &   1.872 \\
    1060A\_393[4] & 10 36 35.60 & -27 08 56.20 & 0.09 & 0.08 & 0.12 & 0.04 & 0.00 & 0.29 &   0.754 \\
    1060A\_393[6] & 10 36 26.76 & -27 23 24.80 & 0.29 & 0.20 & 0.34 & 0.54 & 0.00 & 1.00 &   4.717 \\
    1060A\_393[7] & 10 35 55.70 & -27 14 11.90 & 0.30 & 0.26 & 0.33 & 0.80 & 0.00 & 1.00 &   2.699 \\
   1060A\_393[10] & 10 36 19.30 & -27 28 45.90 & 0.52 & 0.45 & 0.59 & 0.81 & 0.36 & 1.00 &   4.734 \\
   1060A\_393[12] & 10 35 46.79 & -27 38 48.70 & 0.00 & 0.00 & 0.05 & 0.54 & 0.00 & 1.00 &   4.741 \\
   1060A\_393[13] & 10 35 38.27 & -27 31 34.30 & 0.00 & 0.00 & 0.00 & 0.04 & 0.00 & 0.18 &   1.160 \\
   1060A\_393[14] & 10 35 21.68 & -27 23 26.20 & 0.27 & 0.10 & 0.43 & 0.83 & 0.03 & 1.00 &   4.661 \\
   1060A\_393[21] & 10 36 29.47 & -27 45 29.10 & 0.37 & 0.23 & 0.41 & 0.36 & 0.00 & 0.84 &   4.737 \\
   1060A\_393[22] & 10 36 21.38 & -27 46 30.60 & 0.23 & 0.17 & 0.29 & 0.44 & 0.00 & 1.00 &   4.571 \\
   1060A\_393[23] & 10 36 03.81 & -27 55 08.30 & 0.20 & 0.16 & 0.24 & 0.24 & 0.00 & 0.64 &   1.404 \\
   1060A\_393[24] & 10 36 01.81 & -27 41 07.00 & 0.26 & 0.24 & 0.28 & 0.10 & 0.00 & 0.49 &   2.571 \\
   1060A\_393[25] & 10 35 55.21 & -27 45 16.60 & 0.10 & 0.05 & 0.15 & 0.64 & 0.18 & 1.00 &   4.742 \\
   1060A\_393[27] & 10 35 53.88 & -27 22 19.90 & 0.58 & 0.50 & 0.61 & 0.10 & 0.00 & 0.35 &   1.426 \\
   1060A\_393[35] & 10 34 56.05 & -27 38 24.60 & 0.00 & 0.00 & 0.00 & 0.43 & 0.00 & 1.00 &   7.218 \\
   1060A\_393[38] & 10 34 26.69 & -27 30 03.50 & 0.37 & 0.31 & 0.45 & 0.24 & 0.00 & 1.00 &   4.734 \\
   1060A\_393[41] & 10 34 58.90 & -26 52 28.70 & 0.12 & 0.07 & 0.17 & 0.35 & 0.00 & 0.84 &   4.719 \\
   1060A\_393[42] & 10 34 45.84 & -26 57 17.90 & 0.14 & 0.07 & 0.22 & 0.60 & 0.00 & 1.00 &   4.731 \\
   1060A\_393[43] & 10 34 23.74 & -26 59 54.30 & 0.09 & 0.07 & 0.10 & 0.04 & 0.00 & 0.19 &   2.033 \\
\enddata
\tablecomments{$B/T$ values are given as the best fit values and the minimum and maximum values within the 99\% uncertainty interval, as determined by GIM2D \citep{gim2d}. Values with subscript $_{f}$ refer to the fits to images that have been artifically faded to the fiducial redshift of 17858 km/s. The '$\omega$' column lists the statistical weighting factors, calculated by the Discrete Maximum Likelihood method, which corrects for the spectroscopic, morphological and radial incompleteness of the sample. The table contains galaxies down to $m_{R}=18$, not all of which have been used for our analysis. The full table will be published in the electronic version of the Astrophysical Journal and can be provided upon request by the authors. The radial velocities and $R$-band magnitudes are published in \citet{cz2003}.}
\label{tab_catalog}
\end{deluxetable}

Is $B/T$ an adequate proxy for galaxy morphology? The null hypothesis that we test in \S \ref{sec_analysis} is that the star formation-environment relation and the morphology-environment relation are completely interdependent. For testing this hypothesis, it is sufficient to fix a variable that is strongly correlated with morphology. All environmentally-dependent transformation processes suggested in the literature --- including mergers \citep{barnes,bekki98,mihos94} and ram-pressure stripping \citep{gunngott72} --- that could affect star formation also have a strong impact on $B/T$. The bulge fraction is therefore an appropriate variable to use in this analysis.

\subsection{Luminosities and Stellar Masses}
\label{sec_stmass}

We estimate stellar masses in the following way: given a measure of the color of a galaxy, it is possible to reconstruct an approximate mass-to-light ratio and, with a corresponding luminosity measurement, its stellar mass. Infrared magnitudes are best suited for this purpose, because the mass-to-light ratio is most stable in these magnitude bands and least affected by ongoing star formation or dust attenuation. $R$-band magnitudes are available for all galaxies in the sample \citep{cz2003}, and $J$- and $K$-band magnitudes for $\sim80\%$ of them from the 2MASS survey. A rough measure of color that is available for all our galaxies is $D4000$, the strength of the {4000\AA} break. This is defined as
\begin{equation}
D4000=\frac{\int_{4050}^{4250} f_{\lambda} d\lambda}{\int_{3750}^{3950} f_{\lambda} d\lambda}.
\label{eq_d4000}
\end{equation}

We use the GALAXEV spectrophotometric evolution code \citep{bruzualcharlot} to determine a relation between mass-to-light ratios in the $R$-, $J$-, and $K$-band, and $D4000$. Because this relation is dependent on a galaxy's star formation history, we calculate it for three different models: a single-burst model (reflecting a galaxy with an old stellar population), a model with exponentially declining star formation rate (with an e-folding timescale of 2 Gyr, chosen to provide an intermediate scenario between a population dominated by a single burst and one dominated by near-constant star formation), and a model with a constant star formation rate of 1 $M_{\odot}$ yr$^{-1}$ (representative of a field galaxy with on-going star formation). These three models cover a baseline of different, albeit schematic, evolutionary histories and thus allow us to estimate the impact that the choice of model has on the determination of the stellar mass.

From each of the three bands (where available) and each of the three population models, we calculate stellar masses for the galaxies in our sample. We do not know a priori which of the three models best represents a given galaxy, but an accurate population model should yield the same stellar mass estimate from any magnitude band. For this reason, if $J$- and $K$-band photometry are available for a given galaxy, we pick the population model for which the stellar masses calculated from these two bands show the best agreement and adopt the mean of the two masses as the best estimate for the stellar mass of the galaxy, referred to as $M_{*}$. For the $11\%$ of galaxies in the sample that do not have $J$- and $K$-band measurements, we use the $R$-band-derived stellar masses.

To quantify the uncertainty in our stellar mass estimates, we consider the stellar mass values obtained for each galaxy from the three different population models and from the three different filter bands, and calculate their dispersion relative to the best estimate defined above. Typically, the dispersion is on the order of 25-50\%. Because discrepancies among different bands and evolutionary models are systematic rather than statistical, this number is only a rough estimate of the typical uncertainties in the stellar mass.

\subsection{Mean Stellar Age}
\label{sec_msa}
\label{sec_d4000}

We adopt the $D4000$ index, as defined in Eq. \ref{eq_d4000}, as a measure of mean stellar age. In reality, $D4000$ and mean stellar age are not perfectly correlated. To estimate the reliability of $D4000$ as a mean stellar age indicator, we calculate the correlation coefficient between mean stellar age ($MSA$) and $D4000$, $r_{D4000,MSA}$, from a set of GALAXEV population models. We run a variety of models with different star formation histories, consisting of an initial starburst and a variable number of later starbursts, interspersed with episodes of constant star formation, exponentially declining star formation, or quiescence. For all runs, we use the Padova 1994 models with solar metallicity and a Salpeter Initial Mass Function \citep{salpeter}. By running a variety of models with different initial star formation histories and taking one data point at time $t<13.7$ Gyr from each galaxy that has been quiescent for at least 1 Gyr, we find  
\begin{eqnarray}
r_{D4000,MSA}=0.91.
\label{eq_d4000msa}
\end{eqnarray}
Thus D4000 is a very good indicator of the mean stellar age of a galaxy, at least for galaxies that are not currently star forming. In the presence of current or recent star formation, $D4000$ can be biased low. We discuss the impact of this systematic error on our results in \S \ref{sec_errors}.

The star formation history of any galaxy is more complex than can be summarized in a single variable, such as $D4000$. However, for our purpose of establishing whether the star formation gradient results from the recent or current effects of environment as opposed to initial conditions, using $D4000$ as a proxy of star formation history and mean stellar age is adequate.

\section{The Analysis}
\label{sec_analysis}

\subsection{Partial Correlation Coefficients}
\label{sec_partcorrcoeffs}

In our search for trends between environment, star formation, and morphology, we use correlation analyses as our primary tool. The correlation coefficient $r_{xy}$ is a simple and straightforward way of testing a sample for interdependences between any two variables $x$ and $y$. Unlike linear regression slopes, which assume one variable is dependent on the other one, correlation coefficients are symmetric with regard to the variables whose correlation they describe. This is a particular advantage when we want to analyze a data set without inferring a causality beforehand.

However, there are many problems where two variables may be covariate without necessarily being causally related. Both could be dependent on a third, as yet undiscovered variable, or a third variable could be a causal link between the two. If the third variable were somehow fixed at a constant value, no correlation would be observed between the two primary variables. In our analysis, the primary example is the correlation between star formation and environment. Is there a direct causal link between environment and star formation, or can the entire variation in star formation with environment be understood as a consequence of, say, the morphology-environment relation? Determining whether such a third variable exists, what impact it has on the observed correlations, and whether the correlations persist if the variable is held constant, is crucial in examining possible causal connections between environment, star formation, morphology, stellar mass, and mean stellar age.

 To investigate whether there is a dependence of star formation on environment beyond that accounted for by these other variables, we use partial correlation coefficients. Partial correlation coefficients quantify the correlation that two variables, referred to here as $1$ and $2$, would exhibit if a third variable $3$, or even a set of variables $(3,4,...k)$, was held constant, or ``controlled''. 

 Following \citet{kendall} in derivation and notation, the partial correlation coefficient between variables 1 and 2 after partialling out variables 3 through $k$ is
\begin{equation}
r_{12.34...k}=\frac{-C_{12}} {(C_{11}C_{22})^{1/2}} .
\end{equation}
Here, the cofactor $C_{ij}$ is given by
\begin{equation}
C_{ij}=(-1)^{i+j}M_{ij},
\end{equation}
and $M_{ij}$ is the minor of the correlation matrix $(r_{ij})$, i.e., the reduced determinant of the matrix that ensues from discarding row $i$ and column $j$ of $(r_{ij})$. The partial correlation coefficient is therefore only dependent on the total correlation coefficients $r_{ij}$.

The partial correlation coefficients can be tested for significant departures from the null hypothesis in the following way: We follow \citet{kendall} in defining a statistic 
\begin{equation}
\label{eq_z}
Z=\frac{1}{2}\:ln\left(\frac{1+r}{1-r}\right) .
\end{equation}
Unlike the partial correlation coefficient $r$ itself, the probability distribution of $Z$ approximates a normal distribution even for moderate ($>50$) numbers of data points $N$, so that its variance can be used to estimate the significance of the correlation. The variance is approximately given by 
\begin{equation}
var Z = \sigma_{Z}^{2} = \frac{1}{N-1-k} .
\label{eq_zvariance}
\end{equation}

Controlling one variable only reduces the number of degrees of freedom by one. Because our sample contains hundreds of galaxies, our analysis retains discriminatory power even if several variables are partialled out. Thus, we can simultaneously control not only morphology, but also stellar mass and mean stellar age. If the partial correlation coefficient disappears, we can conclude that the star formation gradient can be explained entirely as a result of the simultaneous variations of one or more of these three variables with environment (although the analysis does not tell us immediately whether one of these variables is more important to explaining the star formation gradient than the others, or whether they all contribute to it). If a significant partial correlation coefficient remains, we can conclude that the correlation of star formation with environment is too strong to be explained by environmental variations of either morphology, stellar mass, or mean stellar age, or even all of them together.

Correlation coefficients can be affected strongly by outlying data points, and in such cases, their physical significance is questionable. This is a particular problem for our data, which involve highly non-linear distributions of observables. For example, while most galaxies in the cluster sample have little star formation, and therefore small $EW([OII])$, the few emission-line galaxies may have $EW([OII])$ values many times the typical standard deviation of the entire sample. Ideally, our analysis should be sensitive to correlations both within the bulk of the low-$[OII]$ population and among the outliers to this distribution. 

This problem can be addressed by using the Spearman Rank Correlation Coefficient \citep{gibbons} instead of the Pearson Correlation Coefficient whose definition we have given above. For this purpose, we replace every variable $x$ by its rank $rank(x)$, which is the number of galaxies $i$ with $x<x_{i}$ plus $1/2$ the number of galaxies with $x=x_{i}$. Then, the correlation coefficient is calculated from ranked quantities as described above. Unless explicitly noted otherwise, when we discuss correlations between two variables or provide a correlation coefficient $r_{xy}$, we always refer to rank correlation coefficients. Also note that we use a simplified notation in the subscripts of our correlation coefficients, e.g., $OII$ instead of $EW([OII])$.

\subsection{Completeness-Corrected Correlation Coefficients}
\label{secweighr}

Our sample is incomplete for several reasons: the spectroscopic catalog is incomplete because some galaxies with faint apparent magnitudes and surface brightnesses were not targeted for spectroscopic observations, or observations failed to yield a redshift. The morphological catalog, which contains the results of the bulge/disk decomposition for each galaxy, is incomplete because we only fit galaxies to a limiting magnitude of $m_{R}=18$, and a small number of fits are not successful even for brighter galaxies. Incompleteness introduces an unsystematic weighting into the calculation of the correlation coefficient. For example, if the sample is incomplete at faint magnitudes, and there is a stronger correlation between morphology and star formation for faint galaxies than for bright ones, the correlation coefficient for the incomplete sample will be weighted towards the bright end and artificially weakened. 

To make our results more reproducible, we correct for this incompleteness. The solution is to calculate weighted correlation coefficients, where the weighting factors contain the corrections necessary to account for the sample incompleteness. Our sample spans a range of redshifts, so these weighting factors involve not only the completeness of the spectroscopic and morphological catalogs, but also a volume correction (e.g., a galaxy with $M_{R}=-20$ can be observed in all six clusters, while a galaxy with $M_{R}=-18.5$ is below the detection limit in all but two of our clusters and therefore requires a larger correction). For the calculation of these weighting factors, we use the Discrete Maximum Likelihood method, which we describe in \S \ref{sec_dml}. Although initially developed to calculate luminosity functions, the DML is ideally suited for applications such as these, because it calculates individual weighting factors for each galaxy that comprise all completeness and volume corrections.

To calculate the corrected correlation coefficients, we make use of the definition of the total correlation coefficient as (apart from the sign) the geometric mean of the linear regression slopes of $y$ on $x$ and of $x$ on $y$. To incorporate completeness corrections, we allow for each data point $i$ to be weighted by a factor $\omega_{i}$. This factor, which we calculate in \S \ref{sec_dml}, contains all necessary completeness corrections. The regression slopes are then:
\begin{equation}
b_{yx}=\frac{\sum_{i} (x_{i}-\overline{x})(y_{i}-\overline{y})\omega_{i}}{\sum_{i} (x_{i}-\overline{x})^{2}\omega_{i}} .
\end{equation}
The means $\overline{x}$ and $\overline{y}$ are also computed as weighted means. Analogously, we calculate $b_{xy}$. The correlation coefficient is then given by
\begin{equation}
r_{xy}=\sqrt{b_{yx}b_{xy}}SGN(b_{xy}).
\end{equation}

The sign function, $SGN$, ensures that the correlation coefficient has the same sign as either of the regression coefficients. Since partial correlation coefficients are calculated directly from the total correlation coefficients, this procedure is very straightforward to apply to partial correlation coefficients as well.

In general, these completeness-corrected coefficients do not obey the analytical probability distribution that underlies Eq. \ref{eq_zvariance}; the real probability distribution depends on the choice of $\omega_{i}$ and can be recovered with a Monte Carlo algorithm. In our case, the distribution for the null hypothesis that the expectation value of the correlation coefficient, $<r>$, is zero turns out to be so similar to the analytical prediction for the unweighted case (the standard deviation for the corrected correlation coefficients is $<3\%$ larger) that we can use Eq. \ref{eq_zvariance} to estimate the significance of our results. 

\subsection{The Discrete Maximum Likelihood Method}
\label{sec_dml}

 As we discuss in \S \ref{secweighr}, statistical investigations of a sample of galaxies --- whether by the use of luminosity or other distribution functions, or by multivariate techniques such as in this paper --- require corrections to account for any incompleteness of the sample. 

Maximum Likelihood methods have traditionally been used in studies of galaxy surveys to calculate luminosity functions from samples that are subject to incompleteness and limits in apparent magnitude. The procedure behind these algorithms is to assume a parent distribution function to predict the distribution of galaxies over absolute magnitude and possibly other variables (given the same observational selection effects) and to compare the result to the actual survey. The assumed parent distribution is then adjusted to maximize the probability that the observed sample has been drawn from it. Maximum Likelihood Estimators are very versatile and unbiased by large-scale density inhomogeneities, a particular advantage for field galaxy surveys. However, the functional form and dimensionality of the solution has to be chosen a priori, and the algorithm only solves for a number of pre-selected parameters, discarding information about individual galaxies. Since correlation coefficients are calculated from a set of data points, each of which represents an individual object, this approach is not helpful in our case.

 We have therefore developed the Discrete Maximum Likelihood method specifically to determine completeness and volume corrections for galaxies in parameter spaces of arbitrary dimensionality. Unlike traditional algorithms, the DML associates its free parameters, the weighting factors $\omega_{i}$, not with fixed grid positions in a given parameter space (e.g., with fixed bins along the luminosity axis), but with the individual sampled galaxies. The DML thus combines the advantages of maximum likelihood methods with those of older methods such as the $V/V_{max}$ method, which also associate individual weighting factors with galaxies but that calculate them in ways more susceptible to bias. Our method does not require us to assume a functional form for the distribution function and is, in fact, even independent of the dimensionality of the parameter space in which we are interested. Its disadvantage is that it does not probe regions of parameter space that do not contain galaxies and therefore does not automatically flag regions that could contain large numbers of galaxies outside the survey limits. 

The derivation of the DML is described in detail in \citet{cmz2003}. We therefore only provide an overview here. The principle of the DML is to represent the {\it ansatz} for the distribution function that we wish to recover by the sampled galaxies themselves:

\begin{equation}
\varphi(\vec{x}) = C \sum_{n} \omega_{n} \delta (\vec{x};\vec{x_{n}}) ,
\label{dmlansatz}
\end{equation}

where C is a normalization constant and $\vec{x_{n}}$ is a parameter vector for galaxy $n$ of arbitrary dimensionality. The weighting factors $\omega_{n}$ are free parameters to be determined by the DML algorithm. This {\it ansatz} is similar to that of the {\it C} method \citep{lyndenbell,choloniewski}, but the procedure for solving for the free parameters, $\omega$, is different, with our method retaining the benefits of ML estimators.

Applying this {\it ansatz} to a maximum likelihood approach yields the following iteration formula for the weighting factors $\omega$:
\begin{equation}
\omega_{h}=\left( \sum_{i} \frac{f(\vec{x_{h}}\mid \vec{F_{i}})}{\sum_{g}\omega_{g}f(\vec{x_{g}}\mid \vec{F_{i}})} \right)^{-1} .
\end{equation}
Here, $f(\vec{x_{h}}\mid \vec{F_{i}})$ is the completeness function of the survey --- more specifically, it is the probability that a galaxy with the physical characteristics (e.g., luminosity, rest-frame surface brightness) of galaxy $h$ would have been sampled by our survey if it were in the field (i.e., at the redshift and celestial coordinates, described by the parameter vector $\vec{F}$) of galaxy $i$. While the DML algorithm is independent of the dimensionality of the parameter space that we want to investigate, the calculation of the completeness function is specific to a given survey. We calculate the completeness function of our survey in \S \ref{s_completenessfunction}.

For the faintest galaxies that we consider in our analysis ($M_{R}=-19.2$), we generally find weighting factors $\omega_{h}<3$. 

\subsection{The Completeness Function}
\label{s_completenessfunction}

In this section, we describe how we calculate the completeness function $f(\vec{x}\mid \vec{F})$ of the sample. The calculation of the completeness function is the only part of the DML algorithm that has to be customized for a given survey. 

We assume that our master catalog of photometric detections in the $R$-band is complete, and we determine the completenesses of the spectroscopic and morphological catalogs relative to it. There are three sources of incompleteness to consider. 

\subsubsection{Spectroscopic Incompleteness}

The first, and most severe, source of incompleteness is {\it spectroscopic incompleteness}, $f_{spec}$. For a discussion of how we calculate the spectroscopic completeness function, see \citet{cz2003}. The completeness function (averaged over all six clusters) is also plotted there as a function of apparent magnitude $m_{R}$ and surface brightness $\mu_{R}$.

A source of uncertainty in the spectroscopic completeness function is the assumption that
\begin{equation}
\label{eq_sfrep}
\frac{N_{spec}}{N_{det}}=\frac{N_{spec,cl}}{N_{det,cl}} .
\end{equation}
Here, $N_{det}$ is the total number of photometrically detected galaxies (at a given magnitude and surface brightness), and $N_{spec}$ is the number of spectroscopically sampled galaxies. The index $cl$ indicates the same quantities only for cluster galaxies. This equation assumes that the theoretical sampling fraction for cluster galaxies, in which we are interested, is the same as the empirical sampling fraction for all galaxies, which we can recover from the data. 

The assumption above is affected by statistical uncertainties. The number of sampled cluster galaxies follows a hypergeometric distribution, and the boundaries of the uncertainty interval are approximately given by
\begin{equation}
N_{spec,cl}=N_{spec}\frac{N_{det,cl}}{N_{det}}\left(1\pm(1-\frac{N_{det,cl}}{N_{det}})\frac{N_{det}-N_{spec}}{N_{det}-1}\right) .
\end{equation}

From this, it follows that the real sampling fraction for cluster galaxies is
\begin{equation}
f_{spec,cl}=\frac{N_{spec}}{N_{det}}\left(1\pm(1-\frac{N_{det,cl}}{N_{det}})\frac{N_{det}-N_{spec}}{N_{det}-1}\right) ,
\end{equation}
and the approximate uncertainty in $f_{spec}$ is
\begin{equation}
\label{eq_werror}
\Delta f_{spec}=f_{spec} \left(1-\frac{N_{det,cl}}{N_{det}}\right)\frac{N_{det}-N_{spec}}{N_{det}-1} .
\end{equation}

If we apply our assumption in Eq. \ref{eq_sfrep} to this formula and substitute $\frac{N_{spec,cl}}{N_{spec}}$ for $\frac{N_{det,cl}}{N_{det}}$, we obtain a first-order estimate for the uncertainties arising from this assumption. A stricter treatment should avoid this assumption altogether, but for our purposes, this procedure is sufficient, because, as we show in the next paragraph, the errors are negligible anyway.

Because of the various processing steps to which we subject the completeness function, the best way to estimate the impact of these statistical uncertainties on the final weighting factors $\omega$ is by propagating them through the DML with a Monte Carlo approach. At this point, our intention is only to investigate whether the errors are small enough that they can be neglected safely. For that purpose, we calculate two sets of $\omega$, one using our best-estimate completeness function, and one for which we apply the uncertainty defined by Eq. \ref{eq_werror}  to the sampling fraction values before calculating $\omega$. We then compare the two sets of $\omega$, and find that the mean uncertainty in $\omega$ is less than 5\% even for the fainest galaxies considered here, and the most extreme outliers vary only by $\sim20\%$. A more thorough analysis could estimate uncertainties on the individual $\omega$ by testing a larger number of Monte Carlo realizations, but for our purposes, it is safe to conclude that statistical uncertainties in $\omega$ are sufficiently small so as not to affect our analysis.

Is the spectroscopic completeness of our survey dependent on variables other than $m_{R}$ and $\mu_{R}$, which we explicitly take into account? A concern with many galaxy surveys is that, because of the limited number of spectroscopic apertures available, regions of high galaxy surface number densities may be undersampled. Our survey is not affected by this problem, because we have carried out multiple spectroscopic exposures of each field to ensure that our success in obtaining spectra does not depend on a galaxy's position within the cluster or its proximity to other targets.  Another concern is that because spectroscopic targets were selected in the $b_{J}$-band, rather than the $R$-band, the completeness of our survey may depend on $b_{J}-R$ color. Although this effect is small, we apply a correction here to compensate for it as described in \citet{cz2003}.

\subsubsection{Morphological Completeness}

The second source of incompleteness of our sample is the morphological catalog, which contains bulge-disk decompositions for the majority of our galaxies. We determine this incompleteness empirically as we did for the spectroscopic completeness function, but use $m_{R}$ as the only variable (as opposed to $m_{R}$ and $\mu_{R}$ for the spectroscopic completeness function). Note that bulge-disk decompositions for the artificially faded galaxy images are only available for cluster galaxies with known redshifts. The morphological completeness is therefore highly correlated with the spectroscopic completeness, and it is necessary to calculate it as a conditional completeness, i.e., only among cluster galaxies with spectroscopic information.

\begin{figure}
\plotone{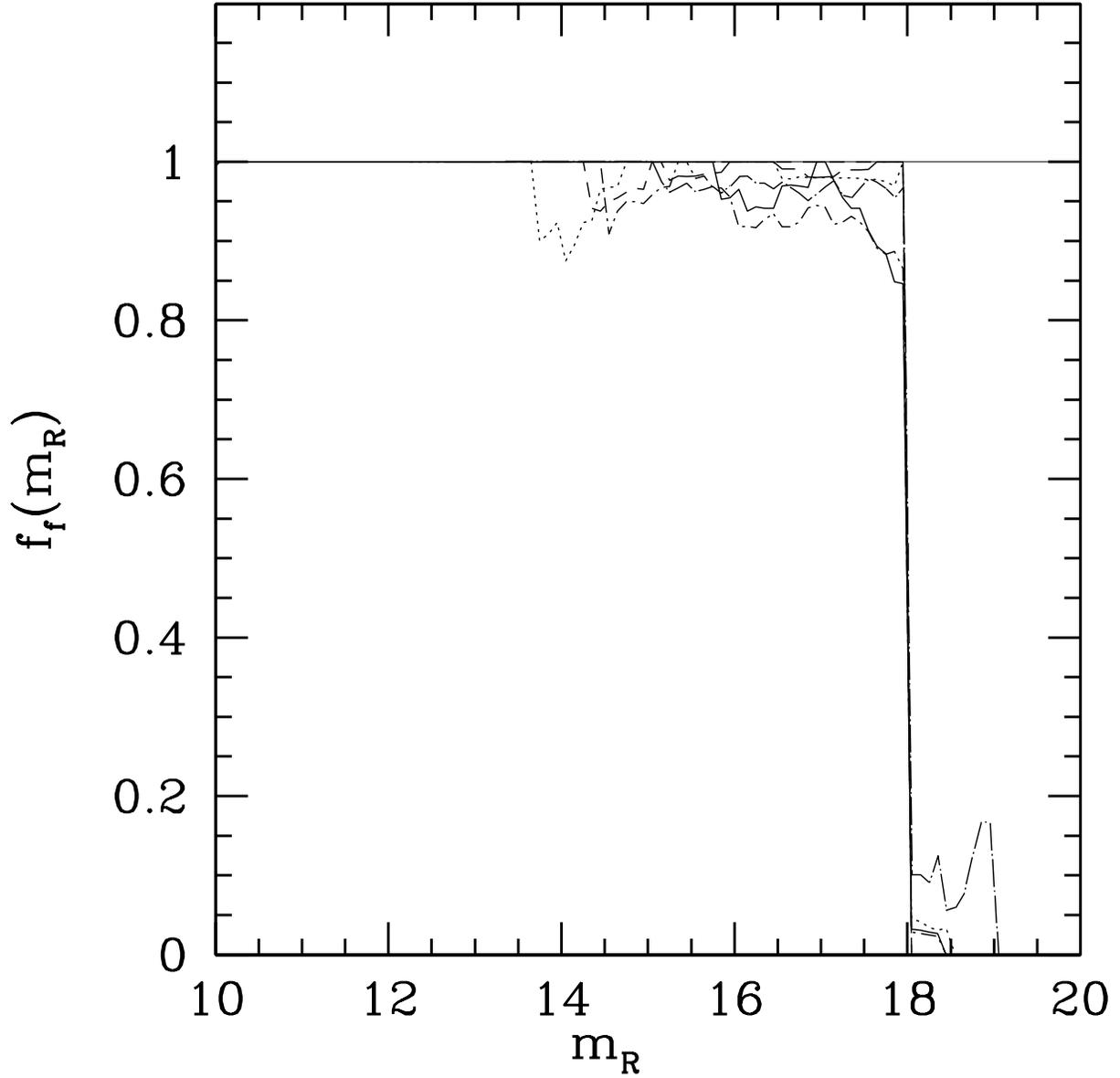}
\caption{Completeness functions of the morphological catalog for all six clusters. Note that completeness is only measured among confirmed cluster members. The morphological catalog has a high level of completeness over all magnitudes brighter than our cutoff limit of $m_{R}=18$. 
}
\label{fig_bdcomp}
\end{figure}

 The completeness functions (only for spectroscopically confirmed cluster members) of all six clusters are shown in Fig. \ref{fig_bdcomp}. The completeness is very high all the way to our cutoff magnitude of $m_{R}=18$. We do not distinguish between incompleteness due to a galaxy not being targeted for a bulge-disk decomposition or due to the bulge-disk decomposition failing.

\subsubsection{Radial Incompleteness}

One selection effect for which we do not correct arises from the spatial boundaries of the survey. In some clusters, the sample includes galaxies at larger projected physical radii than in others. Although the DML allows us to correct for radial sampling incompleteness as well, such a correction would give undue weight (up to $\omega\approx17$) to a small subset of galaxies, severely distorting the probability distribution of the correlation coefficient. Because most of the clusters are sampled to physically comparable radii and radial incompleteness could only introduce a bias indirectly via secondary correlations with cluster global properties, this is not a serious problem for our analysis.

\subsubsection{Summarial Completeness Function}

With contributions from both spectroscopic and morphological selection effects, the completeness function, defined as the probability that a galaxy with the physical parameters (absolute magnitude, rest frame surface brightness) of galaxy $i$ would be observed in the survey if it were located in the field (redshift and spatial coordinates) of galaxy $j$ is
\begin{equation}
f(\vec{x_{i}}\mid \vec{F_{j}}) = f_{morphological}(\vec{x_{i}} \mid \vec{F_{j}},\: spec) f_{spec}(\vec{x_{i}} \mid \vec{F_{j}}) .
\end{equation}

We calculate the summarial completeness function only individually for each galaxy, so we do not plot it here. However, because there is no strong dependence of the morphological completeness function on $m_{R}$ for $m_{R}<18$, the shape of the summarial completeness function is similar to the spectroscopic completeness function shown in \citet{cz2003}.

\section{RESULTS}
\label{sec_results}

\subsection{Morphology-Environment and Star Formation-Environment Relations}
\label{sec_morden}

In this section, we check whether we can detect the morphology-environment relation \citep{oemler74,dressler80} and the star formation-environment relation \citep{lewis,gomez} in our sample. We have examined a range of environmental indices, including local density measures as used by \citet{dressler80}, different scalings of the radial distance from the cluster center, and spatial-kinematic measures such as the Dressler-Shectman $\delta$ statistic \citep{dresslershectman}. All of these indices are significantly correlated with morphology and star formation in our sample. 

It is currently not known whether the properties of galaxies in clusters are primarily a function of their distance from the cluster center \citep{whitmoregilmore91,whitmoretal93} or of local density \citep{dressler80,lewis,balogh04}. Because radius and local density are strongly correlated in our sample, we cannot distinguish between these two possibilities. In this section, however, we are interested simply in testing whether there are significant correlations of morphology and star formation with generic environment.  We select projected radius as our environment index only because local density measurements are typically affected by much larger statistical errors, which compromise our aim of finding an environmental variable that is well-correlated with galaxy properties.

 More specifically, we use the projected physical distance of a galaxy from the cluster centroid (as defined in Table \ref{tabclusters}), scaled by the inverse of the cluster velocity dispersion $\sigma$:
\begin{equation}
R_{\sigma}=R_{phys}\:\sigma^{-1}\:800\:km\:s^{-1} .
\end{equation}
The constant of 800 km s$^{-1}$ is typical of rich clusters \citep{zab90}. The scaling by $\sigma^{-1}$ accounts for the fact that the physical length scales of virialized systems vary as $\sigma$ \citep{girardi}. 

\begin{figure}
\plotone{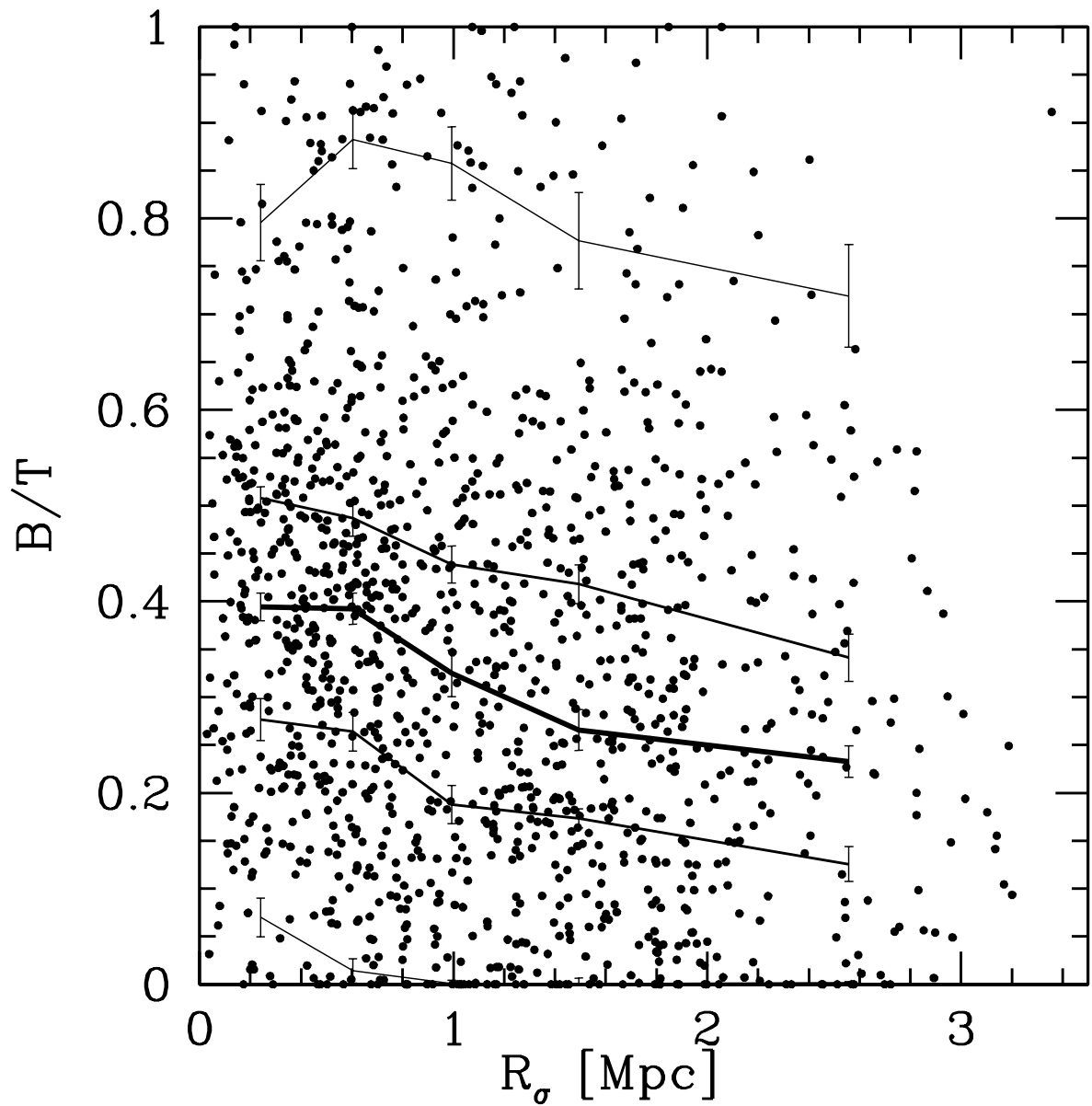}
\caption{Morphology-Environment Relation. The figure shows the (0.0485,0.315,0.5,0.685,0.9515) quantiles of the $B/T$ distribution as a function of cluster-centric radius, $R_{\sigma}$. The median is represented by the bold line. There is a significant correlation of $B/T$ with radius. The whole dynamic range of the median $B/T$ from low to high density environments is $\sim0.2$ to $0.4$.}
\label{fig_morden}
\end{figure}

Fig. \ref{fig_morden} shows the morphology-environment relation for our sample of 1281 cluster galaxies as the median (as well as 0.05, 0.315, 0.685, and 0.95) quantiles of the galaxy distribution over $B/T$ in several bins in $R_{\sigma}$. The magnitude limit for this and for all subsequent analyses is $M_{R}\leq-19.2$. There is a clear correlation of morphology with environment, even though the shift in the median $B/T$ is relatively small compared to the typical range of $B/T$ at each radius. Standard errors on the quantiles are given according to \citet{kendall} as 
\begin{equation}
SE=\sqrt{\frac{p(1-p)}{nf_{1}^{2}}} ,
\end{equation}
where $p$ is the desired quantile, $n$ is the sample size, and $f_{1}$ the frequency of data points per unit interval of the dependent variable near the quantile.

The total rank correlation coefficient for the correlation shown in Fig. \ref{fig_morden} is 
\begin{eqnarray*}
r_{B/T,R_{\sigma}}=-0.215\:(Z=7.8\sigma).
\end{eqnarray*}

\begin{figure}
\plotone{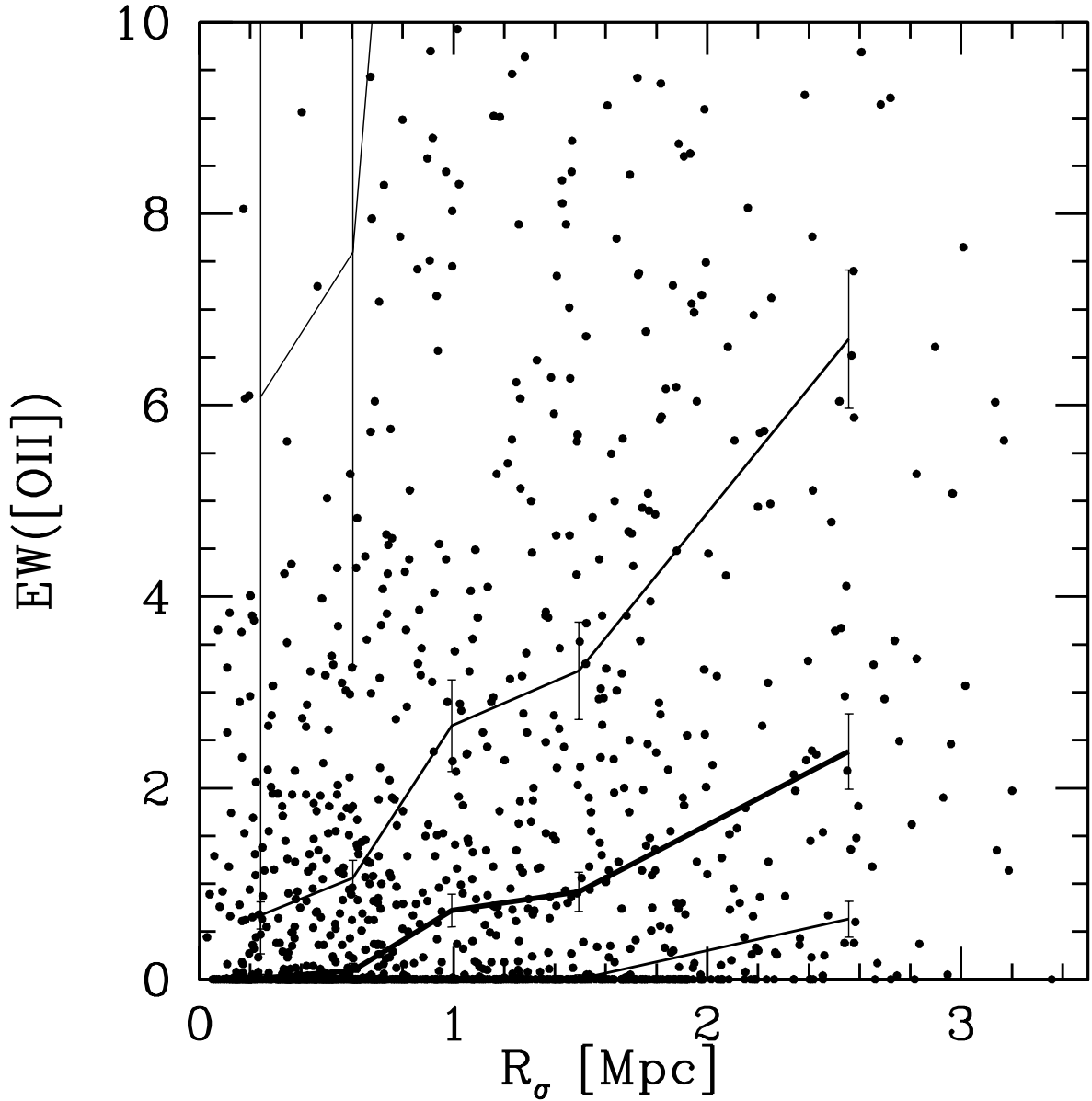}
\caption{Star Formation-Environment Relation. The figure shows the (0.0485,0.315,0.5,0.685,0.9515) quantiles of the $EW([OII])$ distribution as a function of cluster-centric radius, $R_{\sigma}$ (the 0.05-quantile coincides with the $EW([OII])=0$ axis and is therefore not visible in the plot). The median is represented by the bold line. There is a significant correlation of $EW([OII])$ with radius. }
\label{fig_oiiden}
\end{figure}

In analogy to Fig. \ref{fig_morden}, Fig. \ref{fig_oiiden} shows star formation (as measured by $EW([OII])$) as a function of $R_{\sigma}$. There is a highly significant gradient in star formation, with the median $EW([OII])$ in the cluster center being smaller by several {\AA} than at the outskirts. For small radii ($R_{\sigma}<0.5$ Mpc), the median $EW([OII])$ is almost zero. The total correlation coefficient is 
\begin{eqnarray*}
r_{OII,R_{\sigma}}=0.295\:(Z=10.9\sigma) .
\end{eqnarray*}
 Note that our radial sampling limit only extends to $\sim1-2$ virial radii, and therefore does not cover the outer limit at which the star formation rate is claimed to reach the level of the field galaxy population \citep{gomez,lewis}. 

\subsection{Residual Star Formation Versus Environment}
\label{sec_residual}

We have verified above that our sample exhibits the morphology-environment relation as well as the star formation-environment relation.  Now we determine the residual correlation of ongoing star formation with environment while controlling all other variables (morphology, stellar mass, mean stellar age). This approach will show whether the star formation rate varies as a function of its environment even if morphology, stellar mass, and mean stellar age are fixed.

\subsubsection{Residual Correlation of Current Star Formation with Environment}
\label{sec_residualcurrent}

We calculate the partial rank correlation coefficient between $EW([OII])$ and $R_{\sigma}$ while controlling the three variables $B/T$, $D4000$, and $M_{*}$ for all galaxies with $M_{R}\leq-19.2$. With 1281 galaxies, the partial correlation coefficient is
\begin{eqnarray*}
r_{OII,R_{\sigma}.B/T,M_{*},D4000}=0.221\:(Z=8.0\sigma).
\end{eqnarray*}
This highly significant residual correlation means that, even for galaxies with similar morphology (as expressed by $B/T$), stellar mass, and mean stellar age (as expressed by $D4000$), there remains a gradient of current star formation with environment. This result indicates that: 

(1) The star formation gradient is not due simply to the increasing number of long-dead galaxies towards the cluster center. This finding argues that the star formation gradient is not solely the product of initial conditions, but that an environmental mechanism affects the later evolution of galaxies.

(2) The star formation gradient is not just a result of any environmental variations in the galaxy mass function that result in a higher fraction of (star-forming) dwarf galaxies on the cluster outskirts.

(3) The current star formation gradient is not just another aspect of the morphology-environment relation; specifically, the correlation between $EW[OII]$ and environment is too strong to be explained by the relatively weak correlation of $B/T$ with environment. There are several possible explanations:  a) The transformation mechanism affects morphology and star formation differently in different environments (e.g., a galaxy is less likely to re-accrete gas from its tidal tails in the hot, dense cluster core).  b) The transformation mechanism affects star formation on a shorter timescale than morphology, so that recently transformed galaxies show a pronounced star formation gradient, but a weaker morphology-environment relation. c) More than one transformation mechanism is involved, each altering star formation and morphology differently, and each having a different sensitivity to environment.

In this paper, we are unable to discriminate among cases a, b, and c above. However, some relevant insights are provided by \citet{cz2004}, who show that the morphology-environment relation cannot be explained solely by mechanisms --- such as ram-pressure stripping \citep{gunngott72} and strangulation \citep{larson80,balogh00,bekki02} --- in which the disks of galaxies fade over time.  Instead, these authors find that galaxy bulges are systematically brighter toward the cluster center, suggesting that galaxy-galaxy interactions and mergers play a major role in transforming the morphologies of galaxies.  Therefore, if star formation and morphology are driven by same physical process (as in cases a and b above), then ram-pressure stripping and strangulation cannot have a strong effect on the evolution of galaxies.

We note that the current star formation rate inferred from the equivalent width of [OII] may be biased low in a galaxy that has had a burst of star formation in the past few Gyr, due to the significant blue continuum at 3727 \AA. This does not affect our conclusions above for two reasons: First, our principal result is that the observed star formation gradient is {\it too strong} to be explained by correlations of morphology, stellar mass, or mean stellar age with environment. If the real star formation gradient is stronger, our conclusion is a conservative one. Second, an underestimate of the star formation rate in currently star forming galaxies will not only weaken correlations of star formation with environment, but also with morphology, mean stellar age, and stellar mass, so that our procedure of investigating whether these latter three correlations can account for the former remains legitimate.  

To visualize the residual correlation of current star formation with environment, we calculate a residual $\Delta EW([OII])$ for each galaxy with the following procedure: We determine the linear partial regression coefficients $b_{OII,B/T.M_{*},D4000,R_{\sigma}}$, $b_{OII,D4000.M_{*},B/T,R_{\sigma}}$, and $b_{OII,M_{*}.B/T,D4000,R_{\sigma}}$ in rank space \citep{kendall} according to 

\begin{equation}
b_{1,j.2,3,...k}=-\frac{\sigma_{1}}{\sigma_{j}}\frac{C_{1j}}{C_{11}} .
\label{eq_regcoeff}
\end{equation}
The variable $b_{1,j.2,3,...k}$ is the partial regression coefficient of $1$ on $j$, and $\sigma_{j}$ is the square root of the variance of variable $j$. The notation indicates that variables $2,3,...k$ are fixed.

The data can then be represented as
\begin{equation}
<X_{i}>=\overline{X_{i}}+\sum_{j} b_{1,j.2,3,...k}(X_{j}-\overline{X_{j}}) ,
\end{equation}
where $\overline{X_{i}}$ is the mean of variable $i$. In our case, we use this formula to calculate an expectation value for $rank(EW([OII]))$ for each galaxy $i$:
\begin{eqnarray}
<rank(EW([OII]))>\mid_{i}=\overline{rank(EW([OII])_{i})}+\nonumber\\
(rank(B/T_{i})-\overline{rank(B/T)})b_{OII,B/T.D4000.M_{*},R_{\sigma}}+\nonumber\\
(rank(M_{*,i})-\overline{rank(M_{*})})b_{OII,M_{*}.D4000.B/T,R_{\sigma}}+\nonumber\\
(rank(D4000)-\overline{rank(D4000)})b_{OII,D4000.M_{*},B/T,R_{\sigma}} .
\label{eq_pred}
\end{eqnarray}

Note that the indices still represent the ranked variables, not their absolute values. We choose to calculate linear regression coefficients in rank space because the linearity condition is usually much better fulfilled there. 

\begin{figure}
\plotone{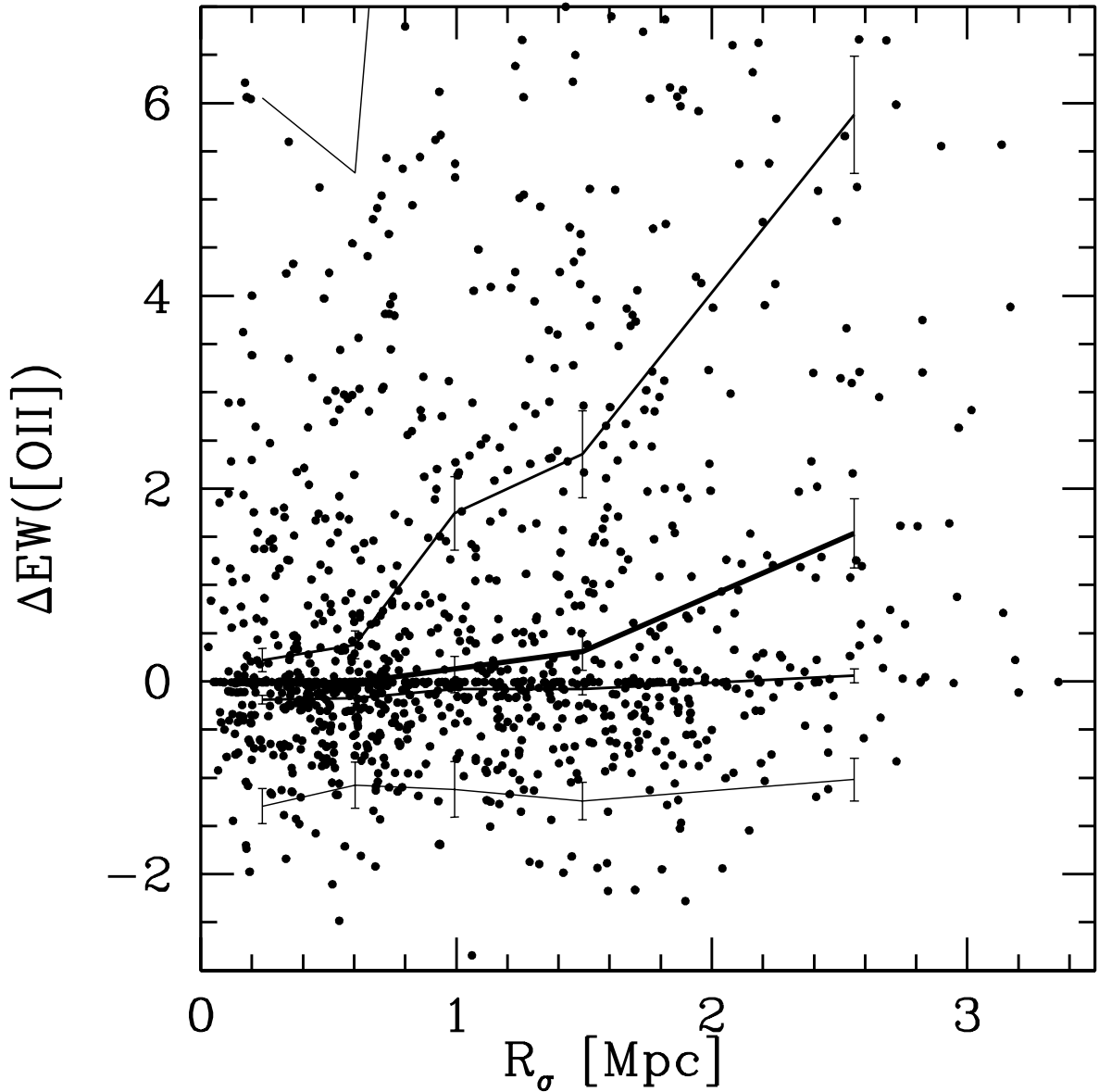}
\caption{$\Delta EW([OII])-R_{\sigma}$ Relation. Dots show the {\it residual} $EW([OII])$ after subtracting the expectation value for a given $B/T$, $D4000$, and $M_{*}$. Lines show the (0.05;0.315;0.5;0.685;0.95) quantiles in each bin, with the median being marked by the bold line. There is a residual correlation, indicating that morphology, stellar mass, and mean stellar age cannot account for the star formation gradient in clusters.}
\label{fig_oiirscle.btd4000stmass}
\end{figure}

We use the observed relation between $EW([OII])$ and $rank(EW([OII])$ to convert the result into an expectation value for $EW([OII])$. This procedure allows us to perform the linear regression in rank space, but to obtain an expectation value in absolute parameter space. We then calculate $\Delta EW([OII])=EW([OII])-<EW([OII])>$, plotting it and the (0.05;0.315;0.5;0.685;0.95) quantiles of the $\Delta EW([OII])$ distribution in Fig. \ref{fig_oiirscle.btd4000stmass}. 

There is a clear residual correlation with environment. Up to a radius of $R_{\sigma}\approx 1$ Mpc, the median residual $EW([OII])$ remains fairly constant, indicating that morphology, stellar mass, and mean stellar age can account for variations of the median $EW([OII])$ over this range. However, for larger radii, the median varies as a function of environment. The (0.685) quantile varies even more steeply with $R_{\sigma}$ than the median, indicating that the properties of strongly star-forming galaxies are particularly correlated with environment.

 The sense of the correlation shows that, for a galaxy with a given morphology, mean stellar age, and stellar mass, current star formation decreases with decreasing cluster-centric radius. 

\subsubsection{Residual Correlation of Recent Star Formation with Environment}
\label{sec_residualrecent}
Our analysis above has placed constraints on the impact of environment on current star formation. Since environmental mechanisms can be differentiated by the timescales on which they operate on a galaxy, it is of considerable interest whether this effect acts not only on current, but on recent star formation as well.

As discussed in \S \ref{sec_sfindices}, we use $EW([H\delta])$ in absorption as an index of recent star formation. We perform this analysis only on a subset of galaxies with no significant [OII] emission. This constraint removes 198 [OII] emitters from the sample, leaving 1083 galaxies. We proceed in the analysis as for $EW([OII])$. There is a very weak, but still detectable correlation; the Spearman correlation coefficient is 
\begin{eqnarray*}
r_{H\delta,R_{\sigma}}=0.084\:(Z=2.8\sigma). 
\end{eqnarray*}

If we remove the effects of $B/T$, $D4000$, and $M_{*}$, the residual correlation coefficient is 
\begin{eqnarray*}
r_{H\delta,R_{\sigma},B/T,M_{*},D4000}=0.064 (Z=2.1\sigma). 
\end{eqnarray*}

\begin{figure}
\plotone{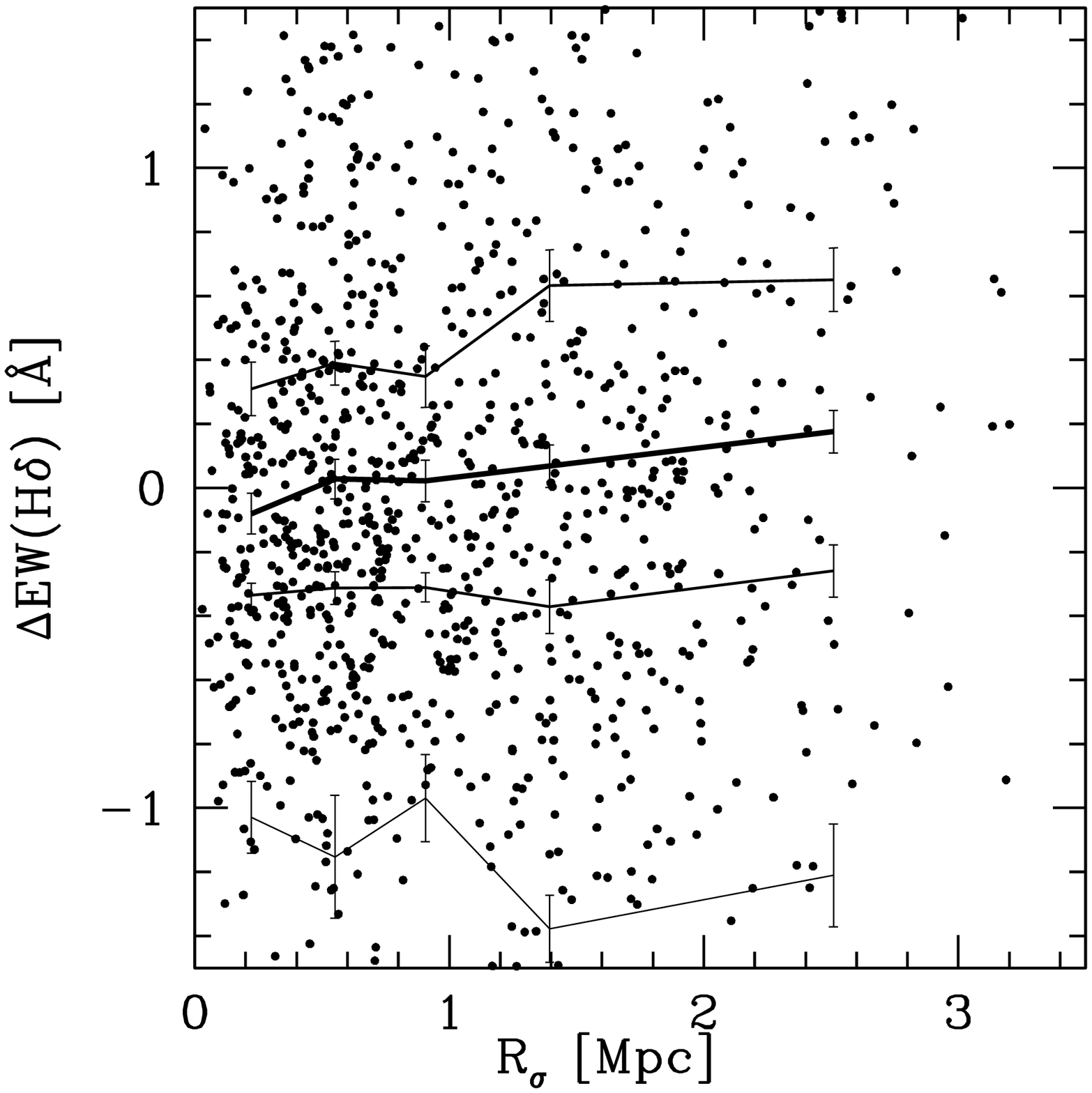}
\caption{{\it Residual} absorption EW[H$\delta$]-$R_{\sigma}$ relation after removing the effects of $B/T$, $D4000$, and $M_{*}$ for non-[OII]-emitting galaxies. Solid lines show the (0.05;0.315;0.5;0.685) quantiles of the $EW([H\delta])$ distribution as a function of cluster-centric radius, $R_{\sigma}$. The 0.95 quantile is off the scale and has very large error bars, and is therefore omitted from this plot. There is a significant residual correlation ($2.1\sigma$) of recent star formation with radius.}
\label{fig_hdelta_residual_stmass}
\end{figure}

Fig. \ref{fig_hdelta_residual_stmass} shows the residual correlation after subtracting the expectation value $<EW(H\delta)>$ for a given ${(B/T,D4000,M_{*})}$ (calculated analogously to Eq. \ref{eq_pred}).

Although this result indicates a significant residual correlation between environment and recent star formation, observational uncertainties in the variables that we are controlling have not been included in this estimate yet, and we briefly comment on them in \S \ref{sec_errors}.

\subsubsection{Effects that Contribute to the Total Star Formation Gradient}
\label{sec_magnitude}

The total (observed) star formation-environment relation is some combination of the morphology-environment, stellar mass-environment, mean stellar age-envrionment, and residual (true) star formation-envrionment relations. Which of these relations drives the total star formation gradient? The absence of an answer to this question has muddled past attempts to draw conclusions about galaxy evolution from the total star formation gradient. In this section, we examine the contributions of each of these relations to the total.

\subsubsubsection{Is the residual star formation gradient a major
component of the total star formation gradient?}

We showed in \S \ref{sec_residual} that there is a residual correlation of star formation with environment when $B/T$, $M_{*}$, and $D4000$ are all held fixed. Here we quantify the contribution of our residual star formation-environment relation to the total star formation gradient. A measure of the actual strength of the residual correlation, compared to the total correlation, is provided by the regression coefficient. The partial regression coefficient in rank space is 
\begin{eqnarray*}
b_{OII,R_{\sigma}.B/T,D4000,M_{*}}=0.192 , 
\end{eqnarray*} 
compared to the total star formation gradient of 
\begin{eqnarray*}
b_{OII,R_{\sigma}}=0.288 . 
\end{eqnarray*} 
In other words, two thirds of the total star formation gradient, on average, is apparent even within a population of galaxies with the same morphology, stellar mass and mean stellar age. Thus, the residual star formation gradient is a major contributor to the total star formation gradient.

The residual star formation gradient in this study is stronger than suggested by earlier studies, which have attempted to model the star formation gradient as a function of the morphology-density relation alone \citep{lewis}. One possible reason for this discrepancy is that Dressler's morphology-density relation \citep{dressler80}, which \citet{lewis} adopt, may be steeper than ours. The Dressler morphology-density relation uses Hubble types, which may themselves be influenced by variations in the star formation rate \citep{koopmann98}.  In such a case, accounting for the effect of morphology removes too much of the star formation gradient (although, to some extent, $B/T$ may also suffer from this problem). Another possible explanation is that \citet{lewis} overestimate the morphology-star formation relation. Their sample does not contain morphological classifications, so they calibrate the morphology-density relation by assuming a perfect correlation between star formation and Hubble types in the field, which is likely to be an overestimate of the true correlation strength. Overestimating either the morphology-star formation or the morphology-environment relation will produce a higher estimate for the contribution of the morphology-density relation to the star formation gradient.

\subsubsubsection{Which other variables contribute significantly to the total star
formation gradient?}

In \S \ref{sec_residual} we showed that controlling $B/T$, $D4000$, and $M_{*}$ simultaneously reduces the correlation coefficient for the star formation-environment relation, indicating that the dependence of these variables on environment does affect the observed star formation gradient. As mentioned above, even the most careful past work has considered only the possible contribution of the morphology gradient to the total star formation gradient.  Is the environment dependence of any of these variables small enough to neglect in discussing the origin of the total star formation gradient?

To answer this question, we examine how strongly the star formation gradient is affected by controlling each of the three variables --- $B/T$, $D4000$, and $M_{*}$ --- {\it individually}. Specifically, we calculate a partial correlation coefficient for the star formation-environment relation while holding only one variable fixed at a time.

First, controlling $B/T$ has the strongest effect on the star formation gradient, indicating that the morphology-environment relation does contribute significantly to the total star formation gradient with environment. Controlling only for $B/T$, we find $r_{OII,R_{\sigma}.B/T}=0.228\:(Z=8.3\sigma)$.

Second, controlling $D4000$ also removes some of the star formation gradient: $r_{OII,R_{\sigma}.D4000}=0.260\:(Z=9.5\sigma)$. However, the impact is rather weak, compared to that of controlling $B/T$, showing that the mean stellar age gradient is not an important contributor to the current star formation gradient. Furthermore, because $D4000$ and $B/T$ are correlated, studies of the star formation-environment relation that control only $B/T$ are automatically accounting for most of the mean stellar age gradient and are therefore not seriously biased.

Third, controlling stellar mass has very little effect on the current star formation gradient. If we control only $M_{*}$, the correlation is reduced from the total $r_{OII,R_{\sigma}}=0.295\:(Z=10.9\sigma)$ to the partial $r_{OII,R_{\sigma}.M_{*}}=0.275\:(Z=10.1\sigma)$. This is because there is only a weak (although significant) segregation of stellar mass with environment $(r_{M_{*},R_{\sigma}}=-.125)$. In this context, we point out that, although there is a segregation of stellar mass with environment, we do not find a significant segregation of $R$-band luminosity with environment. Possibly, the gradient in color is masking the gradient in stellar mass. This result must be kept in mind when searching for luminosity segregation in clusters: even a negative result does not imply an absence of stellar mass segregation.

We conclude that the contribution of the morphology-environment relation to the star formation gradient is far greater than the contributions of the mean stellar age and stellar mass gradients.

\subsection{Effect of Errors and Uncertainties on the Residual Correlations}
\label{sec_errors}

In this section, we examine how strongly measurement uncertainties in our indices of morphology, stellar mass, and mean stellar age affect the residual correlation coefficients. Statistical uncertainties in a variable weaken any total correlations with this variable. This compromises our efficiency at removing the effect of the variable in the calculation of a residual correlation coefficient and will generally cause us to overpredict the significance of a residual correlation coefficient.

We have seen above that the morphology, stellar mass, and mean stellar age do not vary strongly enough with environment and/or star formation to account for the observed star formation gradient with environment. Is that only because our measurements of these variables are affected by large errors? To investigate this possibility, we have to quantify the impact of measurement uncertainties on the residual correlation coefficients. We focus on $B/T$ first, because, for faint galaxies, errors in $B/T$ are of the same order of magnitude as $B/T$ itself.

How much are total and partial correlation coefficients affected by observational uncertainties in one variable? Let us assume that $x$ and $y$ are the ``true'' variables, unaffected by observational uncertainties (e.g., that $y$ is the true bulge fraction), and that $y^{\prime}$ is the observed variable (i.e., the observed $B/T$). For simplicity, we neglect errors in $x$. The partial correlation coefficient when controlling for only one variable can be written out as \citep{kendall}
\begin{equation}
r_{xy^{\prime}.y}=\frac{r_{xy^{\prime}}-r_{xy}r_{yy^{\prime}}}{\sqrt{(1-r^{2}_{xy})(1-r^{2}_{yy^{\prime}})}} .
\end{equation} 

Algebraic transformation yields the ``ideal'' correlation coefficient between the variables that are unaffected by observational uncertainties:
\begin{equation}
r_{xy}=\frac{r_{xy^{\prime}}}{r_{yy^{\prime}}}-\frac{r_{xy^{\prime}.y}}{r_{yy^{\prime}}}\sqrt{(1-r^{2}_{xy})(1-r^{2}_{yy^{\prime}})} .
\label{eq_corrfac}
\end{equation}

If $y^{\prime}$ is merely a noisier version of $y$, then partialing out the ``ideal'' variable $y$ will account for any correlations with $x$ and leave no residual correlation between $x$ and $y^{\prime}$, i.e., $r_{xy^{\prime}.y}=0$. In that case, the ``ideal'' correlation coefficient is
\begin{equation}
r_{xy}=\frac{r_{xy^{\prime}}}{r_{yy^{\prime}}} .
\end{equation}

This relation defines a weakening factor $\xi_{y}\equiv r_{yy^{\prime}}$ that describes how much observational uncertainties in the variable $y$ have reduced a given total correlation coefficient. We determine $\xi_{y}$ empirically by calculating the correlation coefficient between two different realizations --- e.g., two choices of $B/T$ for each galaxy within the GIM2D uncertainty interval --- of the variable $y$. In our data, we find $\xi_{B/T}=0.907\pm0.008$, i.e., all total correlation coefficients that contain the observable $B/T$ are about $10\%$ weaker than they would be if we had a perfect measure of bulge fraction. By scaling all total correlation coefficients that contain $B/T$ by $\xi_{B/T}^{-1}$, we can thus recover the original strength of the correlation coefficient and determine whether the residual correlation coefficient would disappear if observational uncertainties were smaller. 

There is a potential source of systematic errors in the total correlation coefficient between $B/T$ and $EW([OII])$ as well. Aperture bias could introduce a negative correlation, and thus enhance the existing intrinsic correlation between morphology and star formation. The spectrograph used for the spectroscopic survey has a fiber diameter of 3$^{\prime\prime}$, which corresponds to a physical size of 0.7 kpc at the distance of A1060 and 3.4 kpc at the distance of A3266. In galaxies with larger bulges, the fiber captures less light from the disk, and more from the quiescent bulge component. This bias could artificially lower the observed star formation rate.

Given that the observed $r_{OII,B/T}$ is too {\it weak} to explain the star formation gradient, our claim of a residual correlation appears conservative. Nonetheless, we test whether aperture bias is a serious concern. We calculate the correlation coefficient of $EW([OII])$ with angular diameter distance, $D_{A}$, while controlling the bulge absolute magnitude, $M_{bulge}$. Thus, we are comparing whether EW([OII]) varies as a function of distance if we hold the bulge luminosity (which we assume to be correlated with bulge size) fixed. If aperture bias is a problem, then for a given bulge size, $EW([OII])$ should increase with increasing distance as the fiber captures more light from the disk. The correlation coefficient is $r_{OII,D_{A}.M_{bulge}}=+0.038$, ($Z=1.4\sigma$). This effect has almost no impact on the observed correlations: the coefficient for the $EW([OII])-B/T$ correlation after removing $D_{A}$ is $r_{OII,B/T.D_{A}}=-0.442$ at $17\sigma$, practically identical to the total correlation coefficient $r_{OII,B/T}=-0.443$. Therefore, aperture bias does not significantly affect our results.

Our determination of the stellar mass may also be affected by observational uncertainties. The reliability of our stellar mass values can be gauged by comparing the stellar mass estimates based on different filter bands. In our sample, correlation coefficients between our stellar mass values based on $R$-, $J$-, and $K$-band photometry are typically on the order of $r=0.95$. We therefore adopt this as the weakening factor $\xi_{M_{*}}=0.95$, i.e., we assume that all total correlation coefficients containing $M_{*}$ are $\sim 5\%$ weaker than they would be if we had a perfect measure of stellar mass.

With the last of the three variables that we are controlling, $D4000$, there are two sources of uncertainty: one is measurement uncertainty, which we treat in the same way as the measurement uncertainty in $B/T$ and obtain $\xi^{measure}_{B/T}=0.98$. The second source of uncertainty is whether the observable $D4000$, even in the absence of measurement errors, is representative of the variable that we really want to control, the mean stellar age. We have already estimated the correlation coefficient between $D4000$ and the mean stellar age ($MSA$) to be $r_{D4000,MSA}=0.91$ (\S \ref{sec_msa}), and we thus adopt the product of $\xi_{B/T}^{measure}$ and $r_{D4000,MSA}$ as our total weakening factor $\xi_{D4000}=0.89$.

However, galaxies with even moderate current or recent star formation have substantially lower D4000 than the correlation $r_{D4000,MSA}$ above would imply for their mean stellar age. Could this effect generate an artificial residual star formation gradient with environment? If we regard the mean stellar age as the ``ideal'' variable, and $D4000$ as the corresponding observable, then Eq. \ref{eq_corrfac} gives us:
\begin{equation}
r_{OII,MSA}=\frac{r_{OII,D4000}}{r_{D4000,MSA}} - \frac{r_{OII,D4000.MSA}}{r_{D4000,MSA}}\sqrt{(1-r_{OII,MSA}^{2})(1-r_{D4000,MSA}^{2})} .
\label{eq_oiimsa}
\end{equation}

We are concerned about $D4000$ being biased low by current star formation, i.e., about $r_{OII,D4000.MSA}<0$. In that case, ${r_{OII,D4000}}/{r_{D4000,MSA}}$ is a (negative) lower limit on the correlation coefficient $r_{OII,MSA}$. This result, in turn, means that the correlation between mean stellar age and current star formation, $r_{OII,MSA}$, is even weaker than the measured $r_{OII,D4000}$, and we are thus removing too large a mean stellar age gradient from the star formation gradient. Therefore, our claim of a significant residual correlation coefficient is conservative.

What is the numerical effect of all these statistical and systematic uncertainties on our residual correlation coefficient? When we correct the individual total correlation coefficients by the inverse of the weakening factors, we obtain a resulting residual correlation coefficient of
\begin{eqnarray*}
r_{OII,R_{\sigma}.B/T,D4000,M_{*}}=0.192\pm0.037 .
\end{eqnarray*}
Because the residual correlation coefficient after applying our correction factors does not obey the usual probability distribution implied by Eq. \ref{eq_z} anymore (scaling up the total correlation coefficients by $\xi^{-1}$ does not scale up their statistical significances), we have propagated the uncertainties in the individual correlation coefficients into the partial correlation coefficient using a Monte Carlo algorithm. For this reason, we explicitly quote uncertainties for all residual correlation coefficients that we calculate using $\xi$-corrections, instead of providing a level of significance from Eq. \ref{eq_z}. Nonetheless, even with all corrections applied, this result is only slightly smaller than the uncorrected one ($r=0.221$) and is evidence of a significant residual correlation. 

For the observed residual correlation between $EW(H\delta)$ and $R_{\sigma}$, we have repeated this analysis, correcting for the observational uncertainties in $B/T$, $D4000$, and $M_{*}$. We find
\begin{eqnarray*}
r_{H\delta,R_{\sigma}.B/T,D4000,M_{*}}=0.056\pm0.052 .
\end{eqnarray*}

This result suggests, conservatively, that the residual correlation of $EW(H\delta)$ with $R_{\sigma}$ observed earlier at $>2\sigma$ may not be significant.

One last concern: the partial correlation coefficient is calculated from the total correlation coefficients, and not from the properties of the individual galaxies directly. The total correlation coefficients encompass only a limited range of information about the relation between two variables (i.e., they characterize variations of the median, as compared to the variance). In some variables, such as $EW([OII])$ (which is non-negative, but has many outliers at large values), the median does not characterize the galaxy distribution well. Therefore, information is lost in the calculation of the total and, consequently, partial correlation coefficients. The partial correlation coefficient between star formation and projected radius does not test whether individual galaxies show an excess or deficit of star formation over an expectation value based on their morphology, mean stellar age, and luminosity, but only whether the median variations of all these quantities can explain the median variation of star formation with environment.

Is it possible to estimate whether this loss of information affects our conclusions? We do so with regard to the most important of the three variables that we are controlling, $B/T$. We define a new variable, $\delta EW([OII])$, which we obtain by subtracting from the $EW([OII])$ of each individual galaxy the median $EW([OII])$ for galaxies with a similar $B/T$. The variable $\delta EW([OII])$ characterizes the actual excess star formation of a galaxy, relative to the expectation value from its $B/T$. We then apply our correlation analysis to $\delta EW([OII])$ instead of $EW([OII])$. This procedure is more stringent than the standard correlation analysis alone, because it can account better for outlying data points (i.e., a strongly star forming early-type galaxy will influence the correlation coefficient differently than a strongly star forming late-type galaxy). 

We then find a partial correlation coefficient of
\begin{eqnarray*}
r_{\delta OII,R_{\sigma}.B/T,D4000,M_{*}}=0.198\:(Z=7.2\sigma) ,
\end{eqnarray*}
compared to the original estimate of $r_{OII,R_{\sigma}.B/T,D4000,M_{*}}=0.221$. Therefore, the loss of information due to the assumptions that go into the calculation of the correlation coefficient (i.e., linearity, normality) does not compromise our conclusions.

\section{CONCLUSIONS}
\label{sec_conclusions}

We have analyzed the photometric and spectroscopic properties of a sample of galaxies in six nearby clusters. Our primary aim has been to determine whether the star formation-environment relation \citep{lewis,gomez} can be explained as an aspect of the morphology-environment relation \citep{dressler80}, of a mean stellar age gradient, and/or of mass segregation within the cluster, or whether there is evidence that the star formation rates of two galaxies with the same morphology, stellar mass, and mean stellar age vary with environment. Our results are based on a partial rank correlation analysis, which avoids the problems usually associated with binning data, namely, that the finiteness of bin widths may leave residual correlations within each bin.

We find that there is a correlation of current star formation with environment even for galaxies with comparable morphologies, stellar masses, and mean stellar ages. We conclude the following: 

(1) The star formation gradient is not just another aspect of the morphology-environment relation; specifically, the relation between bulge fraction and environment is too shallow to account for the star formation gradient. This result is consistent with some past studies that found that variations of the star formation rate with environment are at least partially independent of certain morphological quantifiers, such as bulge fraction \citep{balogh98}, concentration index \citep{kauffmann2004}, or Hubble type \citep{koopmann98,dressler99,poggianti99,lewis}.

(2) The star formation-environment relation does not arise simply from massive galaxies being biased towards denser environments.

(3) The star formation-environment relation is not solely due to the fact that the centers of clusters are dominated by older galaxies. This result rules out that the star formation gradient was already seeded in initial conditions during the epoch of galaxy formation. The star formation gradient must thus be due in some part to late evolution.

By taking into account variations of morphology, stellar mass, and mean stellar age with environment simultaneously, and nonetheless observing a residual correlation of current star formation with environment, our work constitutes clear statistical evidence for a substantial ongoing effect of environment on galaxy evolution--- a true ``smoking gun.'' 

Although our sample consists of cluster galaxies, it is possible that the environmental mechanisms whose signature we have observed are not specific to clusters. Because of the near-degeneracy of local density with radial distance from the cluster center in our sample, we cannot determine whether the residual star formation gradient is primarily controlled by local density, and may therefore also occur in lower-density environments such as poor groups, or by radial distance, which would suggest that the mechanism or mechanisms responsible for this gradient are found only in the cluster environment.

\acknowledgements

We are grateful to the referee for several helpful comments and questions. Furthermore, we would like to thank Michael Balogh and Mark Swinbank for providing us with a catalog of 2MASS photometry, and to Mark Dickinson, Daniel Eisenstein, Rob Kennicutt, and Bill Tifft for additional comments on the manuscript.

This publication makes use of data products from the Two Micron All Sky Survey, which is a joint project of the University of Massachusetts and the Infrared Processing and Analysis Center/California Institute of Technology, funded by the National Aeronautics and Space Administration and the National Science Foundation.

This research has made use of the NASA/IPAC Extragalactic Database (NED) which is operated by the Jet Propulsion Laboratory, California Institute of Technology, under contract with the National Aeronautics and Space Administration.

A.I.Z. and D.C. acknowledge support from NSF grant \# AST-0206084 and NASA LTSA grant \# NAG5-11108.

\end{document}